\documentclass[12pt]{article}%
\usepackage{graphicx}
\usepackage{amsmath}
\usepackage{amsfonts}
\usepackage{amssymb}
\usepackage{graphicx}
\usepackage{epsfig}
\usepackage[hypertex]{hyperref}%
\providecommand{\U}[1]{\protect\rule{.1in}{.1in}}
\providecommand{\U}[1]{\protect\rule{.1in}{.1in}}
\newtheorem{theorem}{Theorem}

\newtheorem{assumption}[theorem]{Assumption}

\newtheorem{proposition}{Proposition}
\newtheorem{remark}{Remark}

\newenvironment{proof}[1][Proof]{\textbf{#1.} }{\ \rule{0.5em}{0.5em}}
\begin{document}

\title{Energy exchange for homogeneous and isotropic universes with a scalar field
coupled to matter}
\author{Roberto Giamb\`{o}\\Department of Mathematics and Computer Science \\University of Camerino 62032 Camerino (MC) Italy\\and\\John Miritzis,\\Department of Marine Sciences, University of the Aegean,\\University Hill, Mytilene 81100, Greece}
\maketitle

\begin{abstract}
We study the late time evolution of flat and negatively curved
Friedmann-Robertson-Walker (FRW) models with a perfect fluid matter source and
a scalar field arising in the conformal frame of $f(R)$ theories nonminimally
coupled to matter. Under mild assumptions on the potential $V$ we prove that
equilibria corresponding to non-negative local minima for $V$ are
asymptotically stable, as well as horizontal asymptotes approached from above
by $V$. We classify all cases of the flat model where one of the matter
components eventually dominates. In particular for a nondegenerate minimum of
the potential with zero critical value we prove in detail that if $\gamma$,
the parameter of the equation of state is larger than one, then there is a
transfer of energy from the fluid to the scalar field and the later eventually
dominates in a generic way.

\end{abstract}

\section{Introduction}

In cosmological models containing scalar fields the exponential potential
function, $V\left(  \phi\right)  =V_{0}e^{-\lambda\phi}$, is the most popular
not only because of the variety of alternative theories of gravity which
predict exponential potentials, but also due to the fact that this potential
has the nice property that $V^{\prime}\propto V$ which allows for the
introduction of normalized variables according to the formalism of Wainwright
et al \cite{wael}. Among many investigations with exponential potentials, we
mention \cite{cole} and references therein for FRW models; \cite{coibho} for
spatially homogeneous Bianchi cosmologies with an exponential potential;
\cite{hcw} and \cite{copeland} for models containing both a perfect fluid of
ordinary matter and a scalar field with an exponential potential, the
so-called \textquotedblleft scaling\textquotedblright\ cosmologies and
\cite{amen} for nonminimally coupled scalar field; \cite{fay} and references
therein for scalar-tensor theories with exponential potential; \cite{ibol} for
inhomogeneous cosmologies; \cite{coho} for flat and open FRW models and
\cite{aim} for Bianchi type I models with multiple fields containing as a
special case that of a multi-exponential potential studied in the context of
assisted inflation; \cite{cnr} for an elegant mathematical generalization of
multi-exponential potentials; \cite{rspcc} for an investigation of a class of
exponential potentials, allowing exact solutions in the context of dark
energy; \cite{eno} for scalar-tensor phantom cosmologies with exponential
potentials; \cite{ccpr} for the reconstruction of the $f(R)$ theory starting
from the exponential potential in an effort to solve the puzzle of cosmic
acceleration. However, a large class of potentials used in scalar-field
cosmological models have a local minimum. Examples of potentials belonging to
this class are polynomial potentials of the form $V\left(  \phi\right)
=\lambda\phi^{2n},$ generalised and logarithmic potentials $V\left(
\phi\right)  \propto\phi^{n}\left(  \ln\phi\right)  ^{m}$ or $V\left(
\phi\right)  \propto\phi^{n}\exp\left(  -\lambda\phi^{m}\right)  $ studied in
\cite{paba}, potentials used in the study of isotropisation of Bianchi type
models in scalar-tensor theories \cite{miwa} and \cite{fay}, potentials
$V\left(  \phi\right)  \propto Ae^{-\lambda\phi}+Be^{-\kappa\phi}$ in an
effort to avoid eternal acceleration \cite{rspc}, in scalar-tensor
quintessence \cite{peri} (see also \cite{cnp} for potentials reconstructed
from observations), in phantom cosmology \cite{cno}, in double scalar-tensor
cosmologies \cite{cala}. Other important examples include chameleon effective
potentials \cite{khwe} and potentials in conformally related theories of
gravity, for example
\begin{equation}
V\left(  \phi\right)  =\frac{1}{8\alpha}\left(  1-e^{-\sqrt{2/3}\phi}\right)
^{2} \label{rsquared}%
\end{equation}
which arises in the conformal frame of the $R+\alpha R^{2}$ theory
\cite{maed,miri}. Multivalued potentials arising in realistic $f(R)$ gravity
theories were studied in \cite{clno}. Unbounded from below potentials may
drive a flat FRW universe to recollapse \cite{gma}.

Since the nature of the scalar field supposed to cause accelerated expansion
is unknown, it is important to investigate the general properties shared by
all FRW models with a scalar field irrespective of the particular choice of
the potential. Exact solutions for flat FRW models containing only a scalar
field with arbitrary potential were obtained in \cite{reye} (see also
\cite{chri} for Bianchi type I and V models containing ordinary matter).
Nevertheless, the number of papers with mathematically rigorous results is
small. Flat FRW models having an arbitrary potential with a positive lower
bound were studied by Foster \cite{fost}. Generalisation to models including
ordinary matter and scalar fields with potentials having a zero local minimum
was presented in \cite{miri2}. Rendall \cite{rend1} studied the asymptotic
behaviour of homogeneous models with a scalar field having an arbitrary
potential with a positive lower bound and showed that the no-hair theorem
holds for that case (see also \cite{rend2}). Bianchi type I-VIII models were
shown to isotropise in the context of k-essence in \cite{rend3}. The
oscillatory behaviour of scalar fields with a general potential of the form
$V\left(  \phi\right)  =\phi^{2}/2+O\left(  \phi^{3}\right)  $ and ordinary
matter was investigated in \cite{rend4} for flat FRW models. Collapsing models
were built using homogeneous scalar field solutions in \cite{giam1} and the
case where a scalar field is coupled to a perfect fluid was studied in
\cite{giam2}.

The motivation of this investigation comes from a number of physical theories
which predict the presence of a scalar field coupled to matter. For example,
in string theory the dilaton field is generally coupled to matter \cite{gasp}.
Nonminimally coupling occurs also in scalar-tensor theories of gravity
\cite{fuma}, in higher order gravity (HOG) theories \cite{cafa} and in models
of chameleon gravity \cite{wate}. In particular, for HOG theories derived from
Lagrangians of the form
\begin{equation}
L=f\left(  R\right)  \sqrt{-g}+2L_{\mathrm{m}}\left(  \Psi\right)  ,
\label{lagr}%
\end{equation}
it is well known that under the conformal transformation, $\widetilde{g}%
_{\mu\nu}=f^{\prime}\left(  R\right)  g_{\mu\nu}$, the field equations reduce
to the Einstein field equations with a scalar field $\phi$ as an additional
matter source, where
\begin{equation}
\phi=\sqrt{\frac{3}{2}}\ln f^{\prime}\left(  R\right)  . \label{scfi}%
\end{equation}
We do not enter into the discussion about the regularity of the conformal
transformation, or the equivalence issue of the two frames (see for example
\cite{bran}, \cite{maso}, \cite{fara} and references therein). Assuming that
(\ref{scfi}) can be solved for $R$ to obtain a function $R\left(  \phi\right)
,$ the potential of the scalar field is given by
\begin{equation}
V\left(  R\left(  \phi\right)  \right)  =\frac{1}{2\left(  f^{\prime}\right)
^{2}}\left(  Rf^{\prime}-f\right)  , \label{pote}%
\end{equation}
and quadratic gravity with the potential (\ref{rsquared}) is a typical
example. The restrictions on the potential in the papers
\cite{rend1,rend2,rend3,rend4} were used in \cite{macn} to impose conditions
on the function $f\left(  R\right)  $ with corresponding potential
(\ref{pote}). The conformal equivalence can be formally obtained by
conformally transforming the Lagrangian (\ref{lagr}) and the resulting action
becomes \cite{bbpst},
\begin{equation}
\widetilde{S}=\int d^{4}x\sqrt{-\widetilde{g}}\left\{  \widetilde{R}-\left[
\left(  \partial\phi\right)  ^{2}+2V\left(  \phi\right)  \right]
+2e^{-2\sqrt{2/3}\phi}\mathcal{L}_{\mathrm{m}}\left(  e^{-\sqrt{2/3}\phi
}\widetilde{g},\Psi\right)  \right\}  . \label{action}%
\end{equation}
Variation of $\widetilde{S}$ with respect to $\widetilde{g}$ yields the field
equations,
\begin{equation}
\widetilde{G}_{\mu\nu}=T_{\mu\nu}\left(  \widetilde{g},\phi\right)
+\widetilde{T}_{\mu\nu}\left(  \widetilde{g},\Psi\right)  , \label{confm}%
\end{equation}
and variation of $\widetilde{S}$ with respect to $\phi$ yields the equation of
motion of the scalar field,%
\begin{equation}
\widetilde{\square}\phi-\frac{dV}{d\phi}=\frac{1}{\sqrt{6}}e^{-2\sqrt{2/3}%
\phi}T_{\mu}^{\mu}\left(  \widetilde{g},\Psi\right)  . \label{emsf}%
\end{equation}
Note that the Bianchi identities imply that
\begin{equation}
\widetilde{\nabla}^{\mu}\widetilde{T}_{\mu\nu}\left(  \widetilde{g}%
,\Psi\right)  \neq0,\ \ \ \ \widetilde{\nabla}^{\mu}T_{\mu\nu}\left(
\widetilde{g},\phi\right)  \neq0, \label{bian}%
\end{equation}
and therefore there is an energy exchange between the scalar field and
ordinary matter.

In this paper we study the late time evolution of initially expanding flat and
negatively curved FRW models with a scalar field having an arbitrary bounded
from below potential function $V\left(  \phi\right)  $. Ordinary matter is
described by a barotropic fluid with equation of state
\[
p=(\gamma-1)\rho,\ \ \ 0<\gamma\leq2.
\]
The scalar field is nonminimally coupled to matter according to (\ref{action}%
), (\ref{confm}) and (\ref{emsf}). Under general assumptions on the potential
function $V(\phi)$ (see Assumption \ref{assum:V}) we first show in Section
\ref{sec:stable} that stable solutions are related to (possibly degenerate)
local minima of $V$ with non--negative critical value and in cases where $V$
approaches a horizontal asymptote from above. In Section \ref{sec:exchange} we
focus for the sake of simplicity on the flat case and investigate the energy
exchange between the fluid and the scalar field in the case of the exponential
potential and in the case of a nondegenerate local minimum of $V$. In the
former case (Section \ref{sec:exp}) the fluid energy never dominates,
nevertheless the scalar energy always contributes as a nontrivial fraction of
the total energy and in some cases it totally dominates over the fluid, but
stable equilibria may also appear corresponding to scaling solutions,
similarly to the uncoupled case treated in \cite{copeland}. Concerning the
case of nondegenerate minimum of $V$ (Section \ref{sec:nondeg}) we perform a
careful study of the system and prove in detail that the late time qualitative
behaviour is determined by the sign of the pressure of the fluid. In
particular, the energy exchange between the two matter components is such that
for $\gamma>1$ the scalar field eventually dominates (except possibly for a
particular solution, see Theorem \ref{thm:main}). To our knowledge there is no
rigorous proof in the literature of this result and even for the uncoupled
case studied in \cite[Section 4]{rend4} the same conclusion that we find here,
though reasonable, was only conjectured and no proof was provided.

\section{Flat and negatively curved FRW with an arbitrary potential}

For homogeneous and isotropic spacetimes the field equations (\ref{confm})
reduce to the Friedmann equation,
\begin{equation}
H^{2}+\frac{k}{a^{2}}=\frac{1}{3}\left(  \rho+\frac{1}{2}\dot{\phi}%
^{2}+V\left(  \phi\right)  \right)  , \label{fri1jm}%
\end{equation}
and the Raychaudhuri equation,
\begin{equation}
\dot{H}=-\frac{1}{2}\dot{\phi}^{2}-\frac{\gamma}{2}\rho+\frac{k}{a^{2}},
\label{fri2jm}%
\end{equation}
while the equation of motion of the scalar field (\ref{emsf}), becomes
\begin{equation}
\ddot{\phi}+3H\dot{\phi}+V^{\prime}\left(  \phi\right)  =\frac{4-3\gamma
}{\sqrt{6}}\rho. \label{emsjm}%
\end{equation}
The Bianchi identities yield the conservation equation,
\begin{equation}
\dot{\rho}+3\gamma\rho H=-\frac{4-3\gamma}{\sqrt{6}}\rho\dot{\phi},
\label{conssfjm}%
\end{equation}
(see for example \cite{maso,bbpst}). For simplicity we drop the tilde from all
quantities. We adopt the metric and curvature conventions of \cite{wael}.
$a\left(  t\right)  $ is the scale factor, an overdot denotes differentiation
with respect to time $t,$ $H=\dot{a}/a$ and units have been chosen so that
$c=1=8\pi G$. Here $V\left(  \phi\right)  $ is the potential energy of the
scalar field and $V^{\prime}=dV/d\phi$.

In the following, we are going to study the late time behaviour of solutions
of (\ref{fri2jm})-(\ref{conssfjm}), which are expanding at some initial time
of observation, i.e. $H(0)>0$. For flat, $k=0,$ models the state vector of the
system (\ref{fri2jm})-(\ref{conssfjm}) is $\left(  \phi,\dot{\phi}%
,\rho,H\right)  $, i.e. we have a four-dimensional dynamical system subject to
the constraint (\ref{fri1jm}). Defining $y:=\dot{\phi}\ $and setting
$(4-3\gamma)/\sqrt{6}=:\alpha,$ we write the autonomous system as%
\begin{align}
\dot{\phi}  &  =y,\nonumber\\
\dot{y}  &  =-3Hy-V^{\prime}\left(  \phi\right)  +\alpha\rho,\nonumber\\
\dot{\rho}  &  =-3\gamma\rho H-\alpha\rho y,\label{sys1}\\
\dot{H}  &  =-\frac{1}{2}y^{2}-\frac{\gamma}{2}\rho,\nonumber
\end{align}
subject to the constraint
\begin{equation}
3H^{2}=\rho+\frac{1}{2}y^{2}+V\left(  \phi\right)  . \label{cons1}%
\end{equation}

\begin{remark}
\label{rem:W}The function $W$ defined by
\begin{equation}
W(\phi,y,\rho,H)=H^{2}-\frac{1}{3}\left(  \frac{1}{2}y^{2}+V(\phi
)+\rho\right)  , \label{eq:W}%
\end{equation}
satisfies
\begin{equation}
\dot{W}=-2HW. \label{eq:dotW}%
\end{equation}
By standard arguments in ordinary differential equations theory it follows
that $\mathrm{sgn}(W)$ is invariant under the flow of \eqref{sys1}. We deduce
that solutions with $W$ positive, null, or negative, represent scalar field
cosmologies with $k=-1,0,1$ respectively.
\end{remark}

\begin{remark}
\label{rem:rho}Similar arguments applied to the third of \eqref{sys1} show
that if $\rho>0$ at some initial time $t_{0}$, then $\rho(t)>0$ throughout the
solution. Notice that in scalar field cosmologies in the context of GR, the
right hand sides (rhs) of (\ref{emsjm}) and (\ref{conssfjm}) are zero. The
energy exchange between the two matter components (cf (\ref{bian})), is
reflected to the time derivative of the energy density of the scalar field,
\[
\epsilon=\frac{1}{2}y^{2}+V\left(  \phi\right)  ,
\]
which contains the extra term $\alpha\rho y$. Furthermore,
\begin{equation}
\dot{\epsilon}+\dot{\rho}=-3H(y^{2}+\gamma\rho), \label{eq:den}%
\end{equation}
implies that, for expanding models the total energy $\epsilon+\rho$ of the
system decreases.
\end{remark}

\subsection{Stable equilibria for open spatial topologies}

\label{sec:stable}

The equilibria of \eqref{sys1} are given by $(\phi=\phi_{\ast},y=0,\rho
=0,H=\pm\sqrt{V(\phi_{\ast})/3})$ where $V^{\prime}(\phi_{\ast})=0$, and we
are going to discuss their stability for expanding cosmologies ($H>0$) with
open spatial topology ($k=-1,0$).\footnote{The function $V$ is said to have
a\emph{ degenerate} local minimum at $\phi_{\ast}$ if both $V^{\prime}$ and
$V^{\prime\prime}$ vanish at $\phi_{\ast}$. Moreover, if $y,\rho\rightarrow0$
as $t\rightarrow+\infty$ and $\phi\rightarrow+\infty$, while $\lim
_{t\rightarrow+\infty}H\rightarrow\pm\lim_{\phi\rightarrow+\infty}\sqrt
{V(\phi)/3})$, then we call the asymptotic state $(\phi=+\infty,y=0,\rho
=0,H=\pm\sqrt{V(\infty)/3}),$ stable equilibrium \textquotedblleft at
infinity\textquotedblright, although a more appropriate term should be
\emph{stable configuration}.} Hereafter we are dealing with a quite general
class of potentials which includes, for instance, polynomial functions with
even leading term and exponential--decay functions. Essentially all we need is
that $V(\phi)$ is eventually non-negative as $\phi\rightarrow\pm\infty$ and
has a finite number of critical points:

\begin{assumption}
\label{assum:V} We suppose that $V(\phi)\in\mathcal{C}^{2}(\mathbb{R})$
satisfies the following conditions:

\begin{enumerate}
\item the (possibly empty) set $\{\phi\,:\,V(\phi)<0\}$ is bounded;

\item the (possibly empty) set of critical points of $V(\phi)$ is finite.
\end{enumerate}
\end{assumption}

Critical points of $V$ with negative critical value are not equilibria and
they rather allow for recollapse of the model. Moreover, nondegenerate maximum
points (with non negative critical value) for $V$ are unstable, as can be
easily seen by linearizing system \eqref{sys1} at the corresponding equilibria
and verifying the existence of at least one eigenvalue with positive real
part. If the maximum points for $V$ are degenerate, it can be shown that they
are also unstable. In fact, the function $\epsilon+\rho-V(\phi_{\ast})$ is
indefinite in the intersection of the set $\{W\geq0\}$ with an arbitrary
neighborhood of the equilibrium point and strictly decreasing under the flow
of the system for expanding cosmologies (see eq. \eqref{eq:den}). On the
contrary, it can be proved that equilibria corresponding to local minima of
the potential with non-negative critical value are asymptotically stable. More
precisely, we have the following result.

\begin{proposition}
\label{thm:stab} Let $\phi_{\ast}$ a strict local minimum for $V(\phi)$,
possibly degenerate, with nonnegative critical value. Then, $\mathbf{p}_{\ast
}=(\phi_{\ast},y_{\ast}=0,\rho_{\ast}=0,H_{\ast}=\sqrt{\tfrac{V(\phi_{\ast}%
)}{3}})$ is an asymptotically stable equilibrium point for expanding
cosmologies in the open spatial topologies $k=0$ and $k=-1$.
\end{proposition}

The proof of this Proposition is given in the Appendix.

Of course, we must also consider possible stable configurations corresponding
to a diverging value for the field $\phi$. In view of the Assumptions, this
can happen only when $V(\phi)$ has critical points \textquotedblleft at
infinity\textquotedblright\ with asymptotic value $\ell\geq0$. For simplicity
let us restrict to the case $\phi\rightarrow+\infty$. Then it can be shown
that when $V(\phi)\rightarrow\ell^{-}$, the configurations is unstable. The
proof can be done by introducing a new scalar coordinate $\psi=V(\phi)-\ell$,
and writing the system in terms of $\psi$. In this way the critical point at
infinity becomes a finite local maximum ($\psi=0$).
Arguing similarly as sketched before for local maxima, it is shown that when
$V(\phi)\rightarrow\ell^{-}$, the configuration is unstable. On the other
side, when $V(\phi)\rightarrow\ell^{+}$ (e.g. as it happens for the
exponential potential $V\left(  \phi\right)  =V_{0}e^{-\lambda\phi}$) the
critical point at infinity is asymptotically stable, i.e., $y,\rho
\rightarrow0$, and $\phi\rightarrow+\infty$ as $t\rightarrow+\infty$.
Moreover, since $\lim_{\phi\rightarrow+\infty}V\left(  \phi\right)  =\ell
^{+},$ we conclude from (\ref{cons1}) that also $H\rightarrow\sqrt{\ell^{+}%
/3}$ as $t\rightarrow\infty.$ The proof follows after suitable adaptation of
the arguments used in Proposition 4 in \cite{miri2}.

\section{Transition from the matter to the scalar phase}

\label{sec:exchange}

In the following we study the energy transfer from the perfect fluid to the
scalar field. We are interested to study the late time behaviour near
asymptotically stable equilibrium configurations. We suppose that the initial
data in the basin of attraction of this equilibrium are such that the fluid is
the dominant matter component, i.e.
\[
\rho_{0}>\epsilon_{0},
\]
and we are asking whether there is a time $t_{1}$ such that
\begin{equation}
\epsilon\left(  t\right)  >\rho(t),\qquad\forall t>t_{1}.
\label{eq:transition}%
\end{equation}
This question is closely related to the classification of scalar field
potentials with cosmological scaling solutions \cite{lish}. It is also
relevant to the coincidence problem, that is, why dark energy and matter
appear to have roughly the same energy density today (see \cite{aqtw} and
references therein).

There are two cases to be examined for the limiting value of $V(\phi)$,
regardless of the critical point being finite or \textquotedblleft at
infinity\textquotedblright:

(i) $V(\phi)$ asymptotically tends to a strictly positive limit. Then, it is
easily seen that the transition (\ref{eq:transition}) does always happen,
since the critical value of the potential behaves as an effective cosmological
constant and the energy of the scalar field tends to this value whereas the
energy of the fluid tends to zero.\footnote{In a flat FRW model where the
perfect fluid at early times is that of radiation and later that of matter,
dark energy domination during the radiation epoch should be avoided.
Therefore, the relevant question is \emph{when} the scalar dominance would
start. Unfortunately, qualitative methods as those used in this study are
unable to answer this kind of questions.}

(ii) $V(\phi)$ asymptotically tends to zero. This is the nontrivial case and
from now on we will focus ourselves to this question, studying both the
exponential potential and nondegenerate minima. For simplicity we will
restrict ourselves to the flat, $k=0,$ case, although it seems that the same
results can be stated for the general open spatial topology case \cite{prep}.

\subsection{Exponential potential}

\label{sec:exp}

\begin{figure}[ptb]
\begin{center}
\psfull \epsfig{file=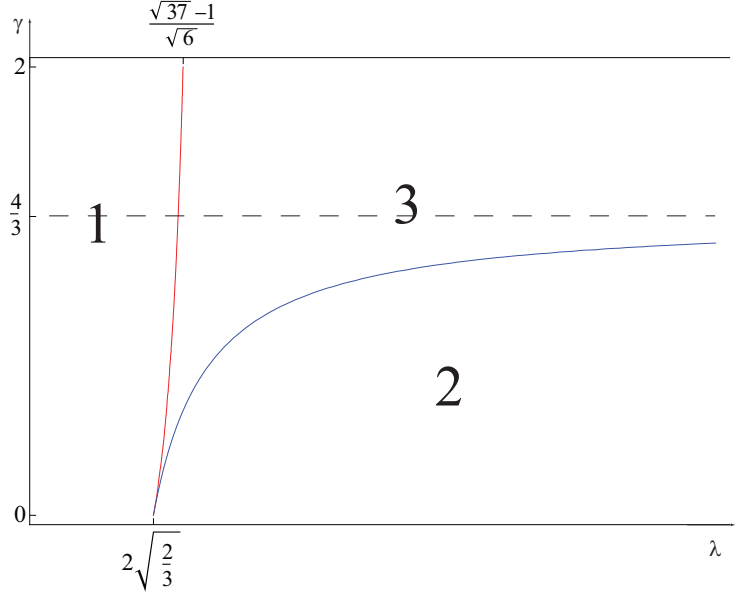, height=8cm}
\end{center}
\caption{Summary of the different stable configurations when $V(\phi
)=V_{0}e^{-\lambda\phi}$. The three regions depicted here corresponds to the
three cases listed on page \pageref{itm:1}.}%
\label{fig:cases}%
\end{figure}

In the case of the exponential potential $V(\phi)=V_{0}e^{-\lambda\phi}$, we
already know that $\phi\rightarrow+\infty$ and $y,\rho\rightarrow0$, (see the
remarks at the end of Section \ref{sec:stable}). To investigate the late time
behaviour of the energy we follow the ideas from \cite{copeland}, writing the
system in the variables
\[
w=\sqrt{\frac{V_{0}}{3}}\frac{e^{-\frac{\lambda}{2}\phi}}{H},\qquad z=\frac
{y}{\sqrt{6}H}%
\]
and $H$, with a new time variable $\tau$ defined by $\mathrm{d}\tau
=3H\mathrm{d}t$:
\begin{align}
\frac{\mathrm{d}w}{\mathrm{d}\tau}  &  =w\left(  -\frac{\lambda}{\sqrt{6}%
}+z^{2}+\frac{\gamma}{2}(1-w^{2}-z^{2})\right)  ,\nonumber\\
\frac{\mathrm{d}z}{\mathrm{d}\tau}  &  =z\left(  -1+z^{2}+\frac{\gamma}%
{2}(1-w^{2}-z^{2})\right)  +\frac{1}{\sqrt{6}}\left(  \lambda w^{2}%
+\alpha(1-w^{2}-z^{2})\right)  ,\label{eq:sysexp}\\
\frac{\mathrm{d}H}{\mathrm{d}\tau}  &  =-H\left(  z^{2}+\frac{\gamma}%
{2}(1-w^{2}-z^{2})\right)  ,\nonumber
\end{align}
where we have used the constraint (\ref{cons1}) to eliminate $\rho$. This
system has the nice property that the third equation decouples from the other
two. Obviously the scalar field dominates when $w^{2}+z^{2}\rightarrow1$.
Studying the Jacobian at the equilibria of \eqref{eq:sysexp} one can find that
there are three possible stable configurations, depending on the couple
$(\lambda,\gamma)$. A summary picture is sketched in Figure \ref{fig:cases}.
In particular:

\begin{enumerate}
\item \label{itm:1} $(w,z)=\left(  \sqrt{1-\frac{\lambda^{2}}{6}}%
,\frac{\lambda}{\sqrt{6}}\right)  $ is a scalar field dominated solution and
is the only stable asymptotic configuration when $\gamma<\frac{\lambda
(4-\sqrt{6}\lambda)}{3(\lambda-\sqrt{6})}$;

\item \label{itm:2} $(w,z)=\left(  0,\frac{4-3\gamma}{3(2-\gamma)}\right)  $
is a scaling solutions, where neither the scalar nor the fluid energy
eventually dominates, and is the only stable asymptotic configuration when
$\lambda>2\sqrt{\frac{2}{3}},\gamma<\frac{4(4-\sqrt{6}\lambda)}{3(2-\sqrt
{6}\lambda)}$;

\item \label{itm:3} $(w,z)=\left(  \frac{(16-4\sqrt6\lambda+3\gamma
(-2+\sqrt6\lambda))^{1/2}}{-4+3\gamma+\sqrt6\lambda},\frac{3\gamma}%
{-4+3\gamma+\sqrt6\lambda}\right)  $ is another scaling solution that is the
only stable asymptotic configuration when $(\lambda,\gamma)$ is on the
complementary of the other two regions.
\end{enumerate}

We can therefore conclude that, in the case of the equilibrium
\textquotedblleft at infinity\textquotedblright\ determined by an
exponential--type potential, the scalar energy $\epsilon$ generically remains
as a nonzero fraction of the total energy. In the first case $\epsilon$
totally dominates over the fluid energy $\rho$, in the other cases both
energies give a nontrivial contribution to the total energy (see also
\cite{howa} for scaling solutions). Interestingly enough, these are the same
conclusions of \cite{copeland} in the case of a scalar field non coupled to
matter, although the ranges of the parameters $(\lambda,\gamma)$ are different.

\subsection{Nondegenerate local minima of $V(\phi)$}

\label{sec:nondeg}

Without loss of generality we suppose that $\phi_{\ast}=0$ and therefore, the
potential can be written in a neighborhood of $\phi=0$ as%
\begin{equation}
V(\phi)=\frac{1}{2}\lambda^{2}\phi^{2}+\mathcal{O}(\phi^{3}),\qquad\lambda>0.
\label{eq:V}%
\end{equation}
It must be stressed that this is the most general form of an arbitrary
potential near the equilibrium $\phi_{\ast}$. In fact the first assumption
does not enter in the study of the late time behaviour around a critical point
$\phi_{\ast}\in\mathbb{R}$, because for that situation only the behaviour of
the potential near the point is important and no further assumptions on $V$
are actually needed. For the sake of simplicity we will neglect higher order
terms in the potential $V(\phi)$. Nevertheless, if we take them into account,
it can be shown by lengthy calculations that the results we are going to state
are not affected. As done in the previous case we introduce new variables $w$,
$z$
\begin{equation}
w=\frac{\lambda}{\sqrt{6}}\frac{\phi}{H},\qquad z=\frac{1}{\sqrt{6}}\frac
{y}{H}, \label{eq:wz}%
\end{equation}
and rewrite the system as
\begin{align}
\dot{w}  &  =\lambda z+3Hw\left(  z^{2}+\frac{\gamma}{2}(1-w^{2}%
-z^{2})\right)  ,\nonumber\\
\dot{z}  &  =-\lambda w+3Hz\left(  z^{2}+\frac{\gamma}{2}(1-w^{2}%
-z^{2})\right)  +3H\left(  -z+\frac{\alpha}{\sqrt{6}}(1-w^{2}-z^{2})\right)
,\label{eq:sys3}\\
\dot{H}  &  =-3H^{2}\left(  z^{2}+\frac{\gamma}{2}(1-w^{2}-z^{2})\right)
,\nonumber
\end{align}
which unfortunately does not possess the decoupling property of the
corresponding system \eqref{eq:sysexp}. Furthermore, we cannot use the time
variable $\tau$ unless we lose regularity properties of the system and we
cannot infer about the stability of the equilibria by looking at the Jacobian
of \eqref{eq:sys3}. Actually we will tackle this nontrivial situation
performing a qualitative study of \eqref{eq:sys3} and prove the following result.

\begin{theorem}
\label{thm:main} Let $\phi_{\ast}$ be a nondegenerate minimum of $V(\phi)$
with zero critical value. Consider the solutions of \eqref{sys1} with $k=0$
approaching the (asymptotically stable) equilibrium point $(\phi_{\ast
},y=0,\rho=0,H=0)$. Then if $\gamma<1$, for every such solution the fluid
energy $\rho$ eventually dominates over the scalar energy $\epsilon$, whereas
if $\gamma>1$, $\epsilon$ eventually dominates over $\rho$ in a generic way,
i.e., except at most for a particular solution of the system.
\end{theorem}

\begin{proof}
As said before, suppose without loss of generality that $V(\phi)$ is as in
\eqref{eq:V}, and consider the variables \eqref{eq:wz} that bring the system
in the form \eqref{eq:sys3}. We first define the (positive) function $R$ by
$R^{2}=w^{2}+z^{2}$ (of course initial data such that $R(0)<1$ implies
$R(t)<1,\forall t>0$) and divide the proof into some steps.\smallskip

\emph{(1) If $\gamma<1$ then $R\rightarrow0$ for every solution.}\smallskip

Let us observe that the flow of the system eventually lives in the compact set
$\{(w,z,H)\,:\,w^{2}+z^{2}\leq1,H\in\lbrack0,H_{0}]\}$ and that the $\omega
$--limit point of every solution, is such that $H=0$. Therefore, if
$(w_{0},z_{0},0)$ is a limit point then it is easily seen by \eqref{eq:sys3}
that the circle $w^{2}+z^{2}=w_{0}^{2}+z_{0}^{2}$, $H=0$ is made by limit
points, which means that the solution asymptotically approaches this circle.
Suppose by contradiction the existence of a solution such that $w_{0}%
^{2}+z_{0}^{2}>0$. Then the only possibility is that $w_{0}^{2}+z_{0}^{2}=1$.
Indeed, passing to polar coordinates
\[
w=R\cos\theta,\qquad z=R\sin\theta
\]
we obtain
\begin{align}
&  \dot{\theta}+\lambda=3H\cos\theta\left(  -\sin\theta+\frac{\alpha}{\sqrt
{6}}\frac{1-R^{2}}{R}\right)  ,\nonumber\\
&  \dot{R}=3H(1-R^{2})\left[  R\left(  \frac{\gamma}{2}-\sin^{2}\theta\right)
+\frac{\alpha}{\sqrt{6}}\sin\theta\right] \label{eq:sys4}\\
&  \dot{H}=-3H^{2}\left(  R^{2}\sin^{2}\theta+\frac{\gamma}{2}(1-R^{2}%
)\right)  .\nonumber
\end{align}
If the solution approaches $R=R_{0}>0$ then the first equation says that
$\theta$ asymptotically behaves like $-\lambda t$. This fact implies the
following asymptotic estimates as $t\rightarrow+\infty$ (recall that
$H\rightarrow0^{+}$ monotonically):
\begin{equation}
\int_{0}^{t}3H\left(  \frac{\gamma}{2}-\sin^{2}\theta\right)  \,d\tau\cong\log
a^{3(\gamma-1)/2},\qquad\int_{0}^{t}3H\sin\theta\,d\tau<+\infty
\label{eq:estimates}%
\end{equation}
and therefore dividing both sides of the second equation in \eqref{eq:sys4} by
$(1-R^{2})$ and using $R\rightarrow R_{0}$, the second equation can be
integrated to give
\[
\lim_{t\rightarrow+\infty}a(t)^{3(\gamma-1)R_{0}}=\frac{1+R_{0}}{1-R_{0}}%
\]
which is consistent with $R_{0}>0$ only when $R_{0}=1$. But this fact implies
$\gamma\geq1$ (when $\gamma=1$ the argument cannot be applied, so we must
include this possibility, see Remark \ref{rem:gammais0} below), a
contradiction.\smallskip

\emph{(2) If $\gamma>1$ and $R\rightarrow0$ then $R/H$ is eventually
bounded.}\smallskip

Let us begin assuming $H/R\rightarrow0$. Then, as above, $\theta\cong-\lambda
t$ and estimates \eqref{eq:estimates} hold. Using $R\rightarrow0$ in the third
equation of \eqref{eq:sys4} we see that $H(t)\cong2/3\gamma t$ and therefore
\[
\left(  \frac{a(t)}{a(t_{0})}\right)  \cong\left(  \frac{t}{t_{0}}\right)
^{\tfrac{2}{3\gamma}}.
\]
Using again $R\rightarrow0$ we can neglect superlinear contribution for
$\dot{R}$ in \eqref{eq:sys4}, and plugging in the above estimate we obtain
\[
R(t)\cong\left(  \frac{t}{t_{0}}\right)  ^{1-\tfrac{1}{\gamma}}\left[
R(t_{0})+c\int_{t_{0}}^{t}s^{1-\tfrac{1}{\gamma}}\sin\theta(s)\,ds\right]
\]
for an appropriate constant $c_{1}$. Possibly shifting the initial instant
$t_{0}$ we can assume that the quantity in square brackets above is strictly
positive, which implies that $R(t)\rightarrow+\infty$ if $\gamma>1$, a
contradiction. Then $H/R\not \rightarrow 0$. To show that this implies that
$H/R$ is bounded away from zero, i.e., $R/H$ bounded, we set $S=H/R$ and
consider system \eqref{eq:sys3} in the new variables $R$, $H$, $S$, $X$ and
$Y$, where
\[
w=\frac{\sqrt{6}}{3\alpha}RX,\ \ z=\frac{\sqrt{6}}{3\alpha}RY.
\]
By hypothesis, the $\omega$--limit points of the solution are such that
$R=0,H=0$ and since the $\omega$--limit set is invariant, the triple $(S,X,Y)$
of a limit point must belong to a solution of the system
\begin{align*}
&  \dot{S}=-S^{2}Y,\\
&  \dot{X}=-Y(-\lambda+SX),\\
&  \dot{Y}=X(-\lambda+SX),
\end{align*}
as it can be easily seen writing the system in the new variables and setting
$R=H=0$. If there exists an $\omega$--limit point of the kind
$(R=0,H=0,S=0,X_{0},Y_{0})$ then using the above system we see that every
point such that $R=0,H=0,S=0$, and $X^{2}+Y^{2}=X_{0}^{2}+Y_{0}^{2}$ is an
$\omega$--limit point, that is, the solution curve approaches the circle
$(R=0,H=0,S=0,X,Y)$ with $X^{2}+Y^{2}=X_{0}^{2}+Y_{0}^{2}$, and in particular
$S\rightarrow0$.\smallskip

\smallskip\emph{(3) If $\gamma>1$ there exists at most one solution such that
$R\to0$ while every other solution is such that $R\to1$.}

Assume by contradiction the existence of two distinct solutions $\xi,\eta$
such that $R(t)\rightarrow0$. By the aforesaid $H/R$ is bounded away from zero
(equivalently, $R/H$ is bounded) for both $\xi,\eta$ and therefore we cannot
exploit anymore the above argument to show $\theta\cong-\lambda t$. A change
of coordinates is then needed in order to control the behaviour of $\theta$
for these solutions. In the following we will derive the appropriate change of
variables for a general system of the form
\begin{subequations}
\begin{align}
&  \dot{w}=\lambda z+F_{1}(w,z,H),\label{eq:w}\\
&  \dot{z}=-\lambda w+F_{2}(w,z,H),\label{eq:z}\\
&  \dot{H}=F_{3}(w,z,H), \label{eq:h}%
\end{align}
where $F_{i}(w,z,H)$ are $C^{\infty}$ functions such that
\end{subequations}
\begin{equation}
F_{i}(0,0,0)=0,\quad(F_{i})_{,w}(0,0,0)=0,\quad(F_{i})_{,z}(0,0,0)=0,
\label{eq:cond-F}%
\end{equation}
and $F_{3}$ is a negative function for $H>0$. It is easily seen that
\eqref{eq:sys3} has the above form. Passing to polar coordinates
$w=R\cos\theta,\,z=R\sin\theta$ we obtain
\begin{subequations}
\begin{align}
&  \dot{\theta}+\lambda=\frac{F_{2}(R\cos\theta,R\sin\theta,H)\cos\theta
-F_{1}(R\cos\theta,R\sin\theta,H)\sin\theta}{R},\label{eq:theta}\\
&  \dot{R}=F_{1}(R\cos\theta,R\sin\theta,H)\cos\theta+F_{2}(R\cos\theta
,R\sin\theta,H)\sin\theta,
\end{align}
and \eqref{eq:cond-F} together with \eqref{eq:theta} suggests that the
condition
\end{subequations}
\begin{equation}
F_{i}(0,0,H)=0,\qquad i=1,2. \label{eq:cond}%
\end{equation}
would ensure $\theta\cong-\lambda t$. Since \eqref{eq:cond} does not hold
\emph{a--priori,} and in particular does not hold for the system we are
considering, we look for a coordinate change
\[
w_{1}=w-f(H),\qquad z_{1}=z-g(H),
\]
where $f,g$ are opportune functions of $H$. Writing the system in the new
unknown functions $w_{1},z_{1},H$ and imposing \eqref{eq:cond} for the new
system we obtain
\begin{subequations}
\begin{align}
\lambda g+F_{1}(f,g,H)-f^{\prime}(H)F_{3}(f,g,H)  &  =0,\label{eq:cond1}\\
-\lambda f+F_{2}(f,g,H)-g^{\prime}(H)F_{3}(f,g,H)  &  =0. \label{eq:cond2}%
\end{align}
This amounts to look for a solution of the original system
\eqref{eq:w}--\eqref{eq:h}. Indeed, if $(w(t),z(t),H(t))$ is such a solution
and denoting by $t(H)$ the inverse function of $H(t)$, we see that
\end{subequations}
\[
f(H)=w(t(H)),\,g(H)=z(t(H)),
\]
satisfy \eqref{eq:cond1}--\eqref{eq:cond2}.

Coming back to the original problem, we will use the solution $\eta$ to
perform the coordinate change described above and apply it to the other
solution $\xi$. Notice also that, since $R\rightarrow0$ and $R/H$ is bounded,
it is easy to verify that the conditions
\begin{subequations}
\begin{align}
&  \lim_{H\rightarrow0^{+}}f(H)=\lim_{H\rightarrow0^{+}}%
g(H)=0,\label{eq:cond3}\\
&  Hf^{\prime}(H),\,\,Hg^{\prime}(H)\quad\text{bounded near}\quad H=0,
\label{eq:cond4}%
\end{align}
hold for $f,g$. Passing to polar coordinates, the new $\theta$ asymptotically
behaves like $-\lambda t$, and it can be found with lengthy but
straightforward calculations that the new $R:=w_{1}^{2}+z_{1}^{2}$ satisfies
the equation
\end{subequations}
\begin{multline}
\dot{R}=3HR\left\{  R^{2}\sin\theta^{2}+2gR\sin\theta+g^{2}+\frac{\gamma}%
{2}(1-R^{2}-f^{2}-g^{2}-2fR\cos\theta-2gR\sin\theta)+\right.
\label{eq:R-transl}\\
\left[  (f+Hf^{\prime}(H))\cos\theta+(g+Hg^{\prime}(H))\sin\theta\right]
\left[  R\sin\theta+2g\sin\theta-\frac{\gamma}{2}(R+2f\cos\theta+2g\sin
\theta)\right]  +\\
\left.  -\sin^{2}\theta-\frac{\alpha}{\sqrt{6}}\sin\theta(R+2f\cos
\theta+2g\sin\theta)\right\}  .
\end{multline}
Therefore under the assumption that $R\rightarrow0$ (note that the new $R$
goes to zero if and only if the old $R$ does, since $f,g\rightarrow0$) we
obtain (using also \eqref{eq:cond3}--\eqref{eq:cond4})
\[
\dot{R}\cong3HR\left(  \frac{\gamma}{2}-\sin^{2}\theta\right)  ,
\]
thus, $R\cong a^{3(\gamma-1)/2}$ which is consistent only if $\gamma\leq1$
(again $\gamma=1$ does not allow to perform the above argument and so must be
included, see again Remark \ref{rem:gammais0} below), a contradiction.
Therefore, except $\eta$ every other solution is such that $R$ does not tend
to zero, and a similar argument as in the step (1) ensures that actually
$R\rightarrow1$. This concludes the proof.
\end{proof}

\begin{remark}
\label{rem:gammais0} We note that the argument of the proof relies on
estimates of the type \eqref{eq:estimates} which are consistent when
$\gamma\neq1$. Therefore we stress that the above argument fails to give
information in the case $\gamma=1$ when actually both situations, $\epsilon
\ll\rho$ and $\rho\ll\epsilon$, may generically occur depending on the initial
data of the problem.
\end{remark}

\section{Discussion}

\label{sec:end}

We analysed the late time evolution of flat and negatively curved expanding
FRW models having a scalar field coupled to matter. We proved that, equilibria
corresponding to non--negative local minima of $V$ are asymptotically stable.
In case the minimum is positive, say $V(\phi_{\ast})>0$, the energy density,
$\epsilon$, of the scalar field eventually rules over the energy density of
the fluid, $\rho$, and the asymptotic state has an effective cosmological
constant $V(\phi_{\ast})$. In case the minimum is zero and nondegenerate, then
$\rho$ eventually dominates over $\epsilon$ if $\gamma<1$ and $\epsilon$
dominates over $\rho$ if $\gamma>1$. This result could be interesting in
investigations of cosmological scenarios in which the energy density of the
scalar field mimics the background energy density. For viable dark energy
models, it is necessary that the energy density of the scalar field remains
insignificant during most of the history of the universe and emerges only at
late times to account for the current acceleration of the universe.

It must be also stressed that the energy transfer from the scalar field to the
fluid showed in Theorem \ref{thm:main} is proved \emph{generically}, i.e. even
in the $\gamma>1$ case there is possibly a special solution such that $\rho$
dominates over $\epsilon$, nevertheless this solution is unstable under
perturbation of initial data of the problem. The generic behaviour of
solutions is not a new feature of models involving scalar fields. We mention
for example the well known instability (with respect to bounded variation
initial data) result by Christodoulou \cite{chris} about scalar fields
collapsing to a naked singularity and even in the realm of homogeneous stars
(regularly matched with nonhomogeneous exteriors), genericity of collapsing
solutions ending in a black hole has been discovered to be true both in GR
\cite{giam1} and in HOG gravity \cite{giam2}.

The above results were rigorously proved only assuming that critical points
are finite, and that $V(\phi)$ is eventually non-negative as $\phi
\rightarrow\pm\infty$. It must be remarked that the latter assumption does not
enter in the study of the late time behaviour around a critical point
$\phi_{\ast}\in\mathbb{R}$, because for that situation only the behaviour of
the potential near $\phi_{\ast}$ is important and no growth at infinity
assumptions on $V$ are actually needed. The other assumption plays the role to
ensure that the flat plateau at infinity are unstable (as proved for the
special case of the potential (\ref{rsquared}) in \cite{miri}), whereas if
$V(\phi)\rightarrow\ell$ from above we have an asymptotically stable
configuration. As a representative of the latter class of stable
configurations we have studied the late time behaviour of the energy in the
case of exponential potential, finding results similar to the non coupled case
\cite{copeland}. However, our dynamical system is much harder to deal with,
because the evolution equation for $H$ does not decouple as in \cite{copeland}
and moreover its equilibria are not hyperbolic. When the decay parameter
$\lambda$ is small, where \textquotedblleft small\textquotedblright\ depends
on $\gamma$, see region 1 in Figure \ref{fig:cases}, the scalar field
dominates, whereas other stable configurations correspond to scaling solutions
where both energies nontrivially contribute to the total energy of the system.

One can encounter in the same potential all the situations studied so far, for
instance, one can have a potential exhibiting more than one local minima, or a
local minimum and an exponential decay at infinity. In that case since at
least two asymptotically stable configurations for the system can be reached,
the late time behaviour of the energies and their exchange necessarily depends
on the initial data of the problem.

The cases studied in Section \ref{sec:exchange} of course do not cover all
possible situations, even under Assumption \ref{assum:V}. In particular, one
can take into account degenerate minima for the potential. In this case, it is
not hard to see that the counterpart of systems \eqref{eq:sys3} and
\eqref{eq:sys4} becomes irregular near $w=z=0$, and to infer about
the behaviour of the angle $\theta$, one cannot straightforwardly extend the
techniques used here. Nevertheless, research in this direction tends to
confirm the same generic behaviour showed here for the nondegenerate case
\cite{prep}. Therefore, it seems appropriate to conclude that energy transfer
is completely driven exclusively by the sign of the fluid pressure when the
scalar reaches a \emph{finite} stable equilibrium, whereas for stable
configurations \textquotedblleft at infinity\textquotedblright\ the potential
profile becomes crucial.

Another important question that should be further investigated is the case of
closed cosmologies. We believe that a closed model cannot avoid recollapse,
unless the minimum of the potential is strictly positive. In that case, the
asymptotic state must be de Sitter space.

\section*{Acknowledgements}

We thank the Referees and the Board Member for their
suggestions which helped us to clarify some points and improve the paper.

\section*{Appendix}

\begin{proof}
[Proof of Proposition \ref{thm:stab}]
%\subsection*{Proof of Proposition \ref{thm:stab}}

\label{a1} We will follow the ideas used in \cite{ggm2} in the absent of
ordinary matter. We start the proof by considering the case $V(\phi_{\ast}%
)>0$. Let $\bar{V}>V(\phi_{\ast})$ be a regular value for $V$ such that the
connected component of $V^{-1}(]-\infty,\bar{V}])$ that contains $\phi_{\ast}%
$, is a compact set in $\mathbb{R}$. Let us denote by $A\subseteq\mathbb{R}$
this compact set and define $\Omega$ as
\[
\Omega=\{(\phi,y,\rho,H)\,:\,\phi\in A,\,(\epsilon+\rho)\leq\bar{V},\,\rho
\geq0,\,W(\phi,y,\rho,H)\in\lbrack0,\bar{W}]\},
\]
where $\bar{W}$ is a positive constant. Then $\Omega$ is a compact set because
the following facts hold:

\begin{enumerate}
\item $\Omega$ is a closed set in $\mathbb{R}^{4}$;

\item $V(\phi_{\ast})\leq V(\phi)\leq\bar{V},\,\forall\phi\in A$;

\item $\tfrac{1}{2}y^{2}+V(\phi_{\ast})\leq\tfrac{1}{2}y^{2}+V(\phi)+\rho
\leq\bar{V}$, and therefore $y$ is bounded;

\item $\rho\leq\bar{V}-\tfrac{1}{2}y^{2}-V(\phi)\leq\bar{V}-V(\phi_{\ast})$,
and therefore $\rho$ also is bounded;

\item from \eqref{eq:W} and the above facts, $\tfrac{V(\phi_{\ast})}{3}\leq
H^{2}\leq\bar{W}+\bar{V}$, and so $H$ is bounded too.
\end{enumerate}

Let $\Omega_{+}\subseteq\Omega$ be the connected component of $\Omega$
containing $\mathbf{p}_{\ast}$. It is easy to see that $H$ is positive on
$\Omega_{+}$ and we claim that $\Omega_{+}$ is \emph{positively invariant}
with respect to \eqref{sys1}, i.e. solutions, with initial data in $\Omega
_{+}$, live in $\Omega_{+}$ $\forall t>0$. Indeed, let $\mathbf{x}(t)$ be such
a solution and $\bar{t}=\sup\{t>0\,:\,H(t)>0\}\in\mathbb{R}\cup\{+\infty\}$.
When $t<\bar{t}$, equations \eqref{eq:dotW} and \eqref{eq:den} imply that both
$W$ and $\epsilon+\rho$ decrease. Moreover, it can be proved by contradiction
that
\begin{equation}
\phi(t)\in A,\,\forall t<\bar{t}, \label{eq:A}%
\end{equation}
otherwise there would exist some $t<\bar{t}$ such that $V(\phi(t)>\bar{V}$,
but then
\[
\bar{V}<V(\phi(t))\leq\frac{1}{2}y^{2}(t)+V(\phi(t)+\rho(t)=\epsilon
(t)+\rho(t)\leq\bar{V},
\]
a contradiction. Thus, \eqref{eq:A} holds. But since $W\geq0$ along the flow
(cf Remark \ref{rem:W}), it follows that
\[
H(t)^{2}\geq\frac{1}{3}(y(t)^{2}+V(\phi(t))+\rho(t))\geq\frac{V(\phi(t))}%
{3}\geq\frac{V(\phi_{\ast})}{3}%
\]
where the last inequality follows from \eqref{eq:A}. All in all we have proved
that, as long as $H$ remains positive it is strictly bounded away from zero,
thus, $\bar{t}=+\infty$, and from this fact it is straightforward to conclude
that $\mathbf{x}(t)$ lives in $\Omega_{+}$ $\forall t>0$.

Using the properties of $\Omega_{+}$ proved so far, LaSalle's invariance
theorem \cite{wig} can be applied to the functions $W$ and $(\rho+\epsilon)$
in $\Omega_{+}$, to show that every solution with initial data in $\Omega_{+}$
must be such that $HW\rightarrow0$ and $H(y^{2}+\gamma\rho)\rightarrow0$ as
$t\rightarrow+\infty$. Since $H$ is strictly bounded away from zero in
$\Omega_{+}$, both $W$ and $(y^{2}+\gamma\rho)$ must go to zero, which means
$y\rightarrow0,\,\rho\rightarrow0$ and $H^{2}-\tfrac{1}{3}V(\phi)\rightarrow
0$. Now the fourth of \eqref{sys1}
\begin{equation}
\dot{H}=-\frac{1}{2}(y^{2}+\gamma\rho)-W, \label{eq:dH}%
\end{equation}
implies that $H$ is monotone and therefore admits a limit. This means that
$V(\phi)$ also admits limit, and this limit must be unavoidably $V(\phi_{\ast
})$, otherwise $V^{\prime}(\phi)$ would tend to a nonzero value, and so would
the right hand side in the second equation of \eqref{sys1}, a contradiction.
Therefore the solution approaches the equilibrium point $\mathbf{p}_{\ast}$.

If $V(\phi_{\ast})=0$, the above argument can be easily adapted. In this case
the set $\Omega$ is connected and we choose $\Omega_{+}$ to be its subset
characterized by the property $H\geq0$. The only point in $\Omega_{+}$ with
$H=0$ is exactly the equilibrium point $\mathbf{p}_{\ast}$, and so if
$H(t)\rightarrow0$ the solution is forced to approach the equilibrium since
$H$ is monotone; if by contradiction $H(t)$ had a strictly positive limit, we
could argue as before to find $y\rightarrow0,\,\rho\rightarrow0$ and
$W\rightarrow0$ and so $H$ must necessarily converge to zero.
\end{proof}

\end{document}